\documentclass[aps,prl,groupedaddress,twocolumn,showpacs]{revtex4}
\usepackage{graphicx,psfrag,amsmath,calc,amssymb}

\begin{document}

\title{Vortex-Antivortex Lattice in Ultra-Cold Fermi Gases}
\author{S. S. Botelho}
\author{C. A. R. S{\' a} de Melo}
\affiliation{School of Physics, Georgia Institute of Technology,
             Atlanta Georgia 30332}

\date{\today}

\begin{abstract}
We discuss ultra-cold Fermi gases in two dimensions, which could be
realized in a strongly confining one-dimensional optical lattice.
We obtain the temperature versus effective interaction  
phase diagram for an $s$-wave superfluid and show that, below a certain 
critical temperature $T_c$, spontaneous vortex-antivortex pairs appear for all 
coupling strengths. In addition, we show that
the evolution from weak to strong coupling is smooth, and that the system 
forms a square vortex-antivortex lattice at a lower critical temperature $T_M$.
\end{abstract}

\maketitle


The presence of quantized vortices is a strong 
indication of the existence of a superfluid state.
The recent experiment from the MIT group~\cite{ketterle-05}
marked the first observation of vortices in atomic Fermi gases. 
This result complements previous experiments involving $s$-wave 
ultra-cold Fermi gases~\cite{hulet-03, ketterle-03, jin-04, grimm-04, 
bourdel-04, kinast-04}.
These studies combined~\cite{ketterle-05, hulet-03, ketterle-03, jin-04, 
grimm-04, bourdel-04, kinast-04}
correspond to the experimental realization of the theoretically proposed 
BCS-to-BEC (weak-to-strong coupling) crossover in three dimensional 
continuum $s$-wave superfluids~\cite{leggett-80a, nozieres-85, sademelo-93}.
More recent generalizations~\cite{perali-04} of these earlier theoretical 
results indicate that the critical temperature in a trap geometry is higher
than in the continuum case throughout the BCS-to-BEC evolution.
$S$-wave Fermi systems have also been investigated in one dimensional 
(1D)~\cite{inguscio-03} 
and three-dimensional (3D)~\cite{esslinger-05} 
optical lattices. Furthermore, the effects of dimensionality have 
been analyzed in $p$-wave Fermi~\cite{esslinger-05-b} and 
Bose~\cite{dalibard-05} systems. In the particular case of Bose systems,
the presence of topological defects associated with the phase of the 
order parameter has been detected in a nearly two-dimensional 
geometry \cite{dalibard-05}. These optical lattice 
experiments~\cite{inguscio-03, esslinger-05, esslinger-05-b, dalibard-05} 
are now allowing the exploration of dimensional (1D, 2D, and 3D) and 
correlation effects in interacting Fermi gases.

Given all the recent advances in experimental techniques, 
we discuss the evolution from weak to strong coupling superfluidity 
of confined 2D $s$-wave ultra-cold Fermi gases.
We show that a Berezinksii-Kosterlitz-Thouless (BKT) 
transition~\cite{berezinskii-71, KT-72} occurs at finite temperatures, 
and that the strong coupling limit produces a critical 
temperature $T_{\rm BKT} = 0.5 \epsilon_F$, where $\epsilon_F$ 
is the Fermi energy. In the strong coupling limit of a 2D Fermi gas, 
the superfluid transition is not characterized by Bose-Einstein 
condensation (BEC) as in 3D, but by the BKT transition.
Below $T_{\rm BKT}$, pairs of vortices and antivortices 
appear spontaneously for all couplings, and eventually condense into a 
vortex-antivortex (VA) square lattice as the temperature is lowered further. 
The lattice melting temperature is shown to be $T_M \approx 0.1 \epsilon_F$ 
in the strong coupling limit, and the melting mechanism is controlled 
by dislocations~\cite{KT-72, nelson-79, young-79}.

The appearance of spontaneously generated 
VA bound states and the existence of 
the VA square lattice are major characteristics of the 
2D physics proposed here. This spectacular effect could be measured
via density or velocity sensitive techniques.
Vortices or antivortices could be detected in a density sensitive experiment
without stirring the condensate, but the topological charge
associated with the sense of rotation should be detectable only in
velocity sensitive experiments.

Two-dimensional Fermi systems can be prepared by means of 
a 1D optical lattice, where tunnelling between lattice sites is suppressed 
by a large trapping potential. A typical trapping potential is
$V_{{\rm trap}} = - V_0 \exp[-2 (x^2 + y^2)/w^2] \cos^2 (k_z z)$, 
where $\lambda_z = 2\pi / k_z$ is the wavelength of the light used 
in the laser beam.
We assume that the width $w$ is such that $w \gg \lambda_F$, where 
$\lambda_F = 2\pi/k_F$ is proportional to the interparticle spacing of a 
Fermi gas with Fermi wavevector ${\bf k}_F$. The atom transfer energy 
along the $z$ direction can be estimated using a simple WKB expression leading
to $t_z \approx W_0 \exp \left[ -\pi \sqrt{(V_0 - W_0)/E_r} \right]$,
where $W_0 =  \hbar \omega_0/2 = \sqrt{V_0 E_r/2}$. 
Here, $E_r = \hbar^2 k_z^2/2M$ is the recoil energy of atoms with mass $M$.
Thus, in the 2D limit of $t_z \to 0$, there is no finite
critical temperature for BEC, and the superfluid phase in both the BEC 
and BCS regimes is characterized by VA bound pairs.


To explore the physics described above, we consider a 2D continuum model 
of Fermi atoms of mass $M$ and density $n = k_F^2 / 2\pi$, 
with Hamiltonian ($\hbar=k_B=1$)
\begin{equation}
{\cal H}=\sum_{{\bf k}, \sigma} 
\xi_{\bf k} \psi_{{\bf k} \sigma}^\dagger \psi_{{\bf k} \sigma} + 
\sum_{{\bf k},{\bf k}',{\bf q}} V_{{\bf k}{\bf k}'}
b_{{\bf k}{\bf q}}^\dagger b_{{\bf k}'{\bf q}} ,
\end{equation}
where 
$
b_{{\bf k}{\bf q}} = \psi_{-{\bf k}+{\bf q}/2, \downarrow}
\psi_{{\bf k}+{\bf q}/2, \uparrow}$ 
and
$\xi_{\bf k}= \epsilon_{\bf k}  - \mu$, 
with energy dispersion $\epsilon_{\bf k} = k^2/2m$ and
chemical potential $\mu$. Following the procedure discussed in detail
in Ref.~\cite{duncan-00}, we obtain the following separable
expression for the interparticle potential in $k$-space,
\begin{equation}
\label{potential}
V_{{\bf k}{\bf k}'} = -\lambda 
\Gamma ({\bf k}) \Gamma ({\bf k}') ,
\end{equation}
where $\lambda$ is the interaction strength and,
in the particular case of $s$-wave symmetry ($\ell = 0$), 
$
\Gamma({\bf k}) = 
\left( 1 + k/k_0 \right)^{-1/2},
$
with $R_0 \sim k_0^{-1}$ playing the role of the interaction range.


The partition function $Z$ at temperature $T = \beta^{-1}$ is 
an imaginary-time functional integral with action
$
S=\int_0^\beta d\tau [ \sum_{{\bf k}, \sigma}
\psi_{{\bf k} \sigma}^\dagger(\tau) \partial_\tau 
\psi_{{\bf k} \sigma}(\tau) + {\cal H} ] .
$
Introducing the Hubbard-Stratonovich field $\phi_{\bf q}(\tau)$, 
which couples to $\psi^\dagger\psi^\dagger$, and integrating out 
the fermionic degrees of freedom, we obtain 
\begin{equation}
Z=\int {\cal D}\phi{\cal D}\phi^* \,
\exp(-S_\mathrm{eff}[\phi,\phi^*]) ,
\end{equation}
with the effective action given by
$$
S_{\rm{eff}} = \int_0^\beta d\tau \sum_{\bf k}
\left( \frac {|\phi_{\bf k}(\tau)|^2} {\lambda} + \xi_{\bf k} \right) -
{\bf Tr} \left( \ln \mathbf{G}_{{\bf k},{\bf k}^\prime}^{-1}(\tau) \right) .
$$
The symbol ${\bf Tr}$ denotes the trace over momentum, imaginary time and 
Nambu indices, and $\mathbf{G}_{{\bf k}, {\bf k}^\prime}^{-1}(\tau)$ 
is the (inverse) Nambu matrix,
\begin{equation}
\mathbf{G}_{{\bf k},{\bf k'}}^{-1}(\tau) = 
\left(
\begin{array}{cc}
-(\partial_\tau + \xi_{\bf k})\delta_{{\bf k},{\bf k}^\prime} & 
 \Lambda_{ {\bf k}, {\bf k'}} (\tau) \\[3mm]
\Lambda^*_{ {\bf k'}, {\bf k}} (\tau) &
-(\partial_\tau - \xi_{\bf k}) \delta_{{\bf k},{\bf k}^\prime} 
\end{array}
\right),
\end{equation}
with
$
\Lambda_{{\bf k},{\bf k}^\prime} (\tau) = 
\phi_{{\bf k} - {\bf k}^\prime} (\tau) 
\Gamma ( ({{\bf k} + {\bf k}^\prime})/2) .
$ 

Assuming 
$
\phi_{\bf q}(\tau) = \Delta_0 \delta_{{\bf q}, 0} +
\eta_{\bf q}(\tau)
$
and performing an expansion in $S_{\rm eff}$ to quadratic order 
in $\eta$, one can write the effective action as 
$
S_{\rm eff} = S_0[\Delta_0] + 
S_{\rm fluct}[\eta, \eta^{*}] ,
$
where $S_0[\Delta_0]$ is the saddle point action, and the fluctuation 
action $S_{\rm fluct}[\eta, \eta^{*}]$ can be expressed in terms of 
amplitudes and phases via $\eta(q) \equiv |\eta(q)| e^{i\theta(q)}$.
Integrating out the amplitudes, and Fourier transforming 
to position and imaginary time $r = ({\bf r}, \tau)$, one obtains
$$
S_{\rm fluct} = \frac {1} {2} \int dr \, \big[
\rho_{ij} \, \partial_i \theta(r) \partial_j \theta(r) +
A (\partial_{\tau} \theta)^2 
-i B \partial_{\tau} \theta(r) \big] .
\label{eqn:fluctuation-action}
$$
In this expression,
$$
A(\mu, \Delta_0, T) =  \frac {1} {4 L^2} \sum_{\bf k} 
\left[
\frac {|\Delta_{\bf k}|^2} { E_{\bf k}^3 }
\tanh \left( \frac { E_{\bf k} }  { 2 T}  \right) + 
\frac { \xi_{\bf k}^2 } { E_{\bf k}^2 } Y_{\bf k}
\right] ,
$$
where 
$
Y_{\bf k} = (2T)^{-1} 
{\rm sech}^2 (E_{\bf k} / 2T)
$
is the Yoshida distribution,
and
$$
B(\mu, \Delta_0, T) = \frac {1} {L^2} \sum_{\bf k}
\frac {\xi_{\bf k}} {E_{\bf k}} 
\tanh \left( \frac {E_{\bf k}} {2T} \right).
$$
The quantity $\rho_{ij}$ represents the 
superfluid density tensor
$$
\rho_{ij}(\mu, \Delta_0, T) = \frac {1} {L^2} \sum_{\bf k} \left[
2 n_0 ({\bf k}) \partial_i \partial_j \xi_{\bf k} - 
Y_{\bf k} \partial_i \xi_{\bf k} \partial_j \xi_{\bf k}
\right] ,
$$
where $\partial_i$ denotes the partial 
derivative with respect to $k_i$, and
$$
n_0 ({\bf k}) = \frac {1} {2} \left[
1 - \frac {\xi_{\bf k}} {E_{\bf k}} 
\tanh \left( \frac { E_{\bf k} } {2 T} \right) \right]
$$
is the momentum distribution. 
Notice that $\rho_{xx} = \rho_{yy} \equiv \rho_s$, 
while $\rho_{xy} = \rho_{yx} = 0$.

The decomposition of 
$\theta ({\bf r}, \tau) = \theta_v ({\bf r}) + \theta_s ({\bf r}, \tau)$
into a static vortex part $\theta_v ({\bf r})$ and a 
spin-wave part $\theta_s ({\bf r}, \tau)$
permits us to rewrite
\begin{equation}
S_{\rm fluct} = S_{v} + S_{sw},
\end{equation}
where 
$S_{v} = \frac{1}{2} \int dr \rho_s\left[ \nabla \theta_v (r) \right]^2$,
while 
$
S_{sw} = \frac{1}{2} \int dr 
\left[{ \rho_s \left[ \nabla \theta_{sw} (r) \right]^2
+ A \left[\partial_\tau \theta_{sw} (r) \right]^2
- iB \partial_{\tau}\theta_{sw} (r) } \right].
$
The spin-wave part can be integrated out to give
$
\Omega_{sw} = \sum_{\bf q} 
T\ln \left[ 1 - \exp \left( -  { w_{\bf q} / T } \right) \right] ,
$
where $w ({\bf q}) = c |{\bf q}|$ is the frequency and 
$c = \sqrt {\rho_s/A}$ is the speed of the spin-wave. 
Here, $\Omega_{sw}$ is the spin-wave contribution to the thermodynamic 
potential $\Omega = \Omega_0 + \Omega_{sw} + \Omega_v$, where
$\Omega_0 = T S_0 [\Delta_0]$ and $\Omega_v$ are the saddle-point and
vortex parts, respectively.

The self-consistent equations for $\mu$, $\Delta_0$ and $T_c$ 
can be derived from the effective action $S_{\rm eff} = S_0 + S_{sw} + S_v$ 
as follows.
%
%
The order parameter equation is obtained through
the stationarity condition 
$
[\delta S_{\rm eff} / 
\delta \phi^*_{\bf q}(\tau')]_{\Delta_0} = 0 ,
$
leading to 
\begin{equation}
\label{eqn:op}
\frac {1} {\lambda}  = 
\sum_{\bf k} \frac {|\Gamma({\bf k})|^2} {2 E_{\bf k}}
\tanh \left( \frac {E_{\bf k}} {2 T} \right) ,
\end{equation}
where
$
E_{\bf k} = \sqrt{\xi_{\bf k}^2 + |\Delta_{\bf k}|^2}
$
is the quasiparticle excitation energy, and 
$\Delta_{\bf k} = \Delta_0 \Gamma ({\bf k})$ plays the role
of the order parameter function. 
Elimination of the interaction strength $\lambda$ in favor of the two-body 
bound state energy $E_b$ in vacuum is possible through the relation
\begin{equation}
\frac {1} {\lambda} = 
\sum_{\bf k} \frac { |\Gamma({\bf k})|^2 } 
{2\epsilon_{\bf k} - E_b}.
\end{equation}

The number equation is obtained via 
$
N_p = -\partial\Omega_p / \partial\mu,
$
where 
$
\Omega_p = \Omega_0 + \Omega_{sw} ,
$
leading to 
\begin{equation}
\label{eqn:number}
N_p  =  N_0 + N_{sw}. 
\end{equation}
Here, 
$N_0 = - \partial \Omega_0 / \partial \mu = 2 \sum_{\bf k} n_0 ({\bf k})$, 
and $N_{sw} = -  \partial \Omega_{sw} / \partial \mu$.

\begin{figure}
\begin{center}
\includegraphics[width=4.1cm]{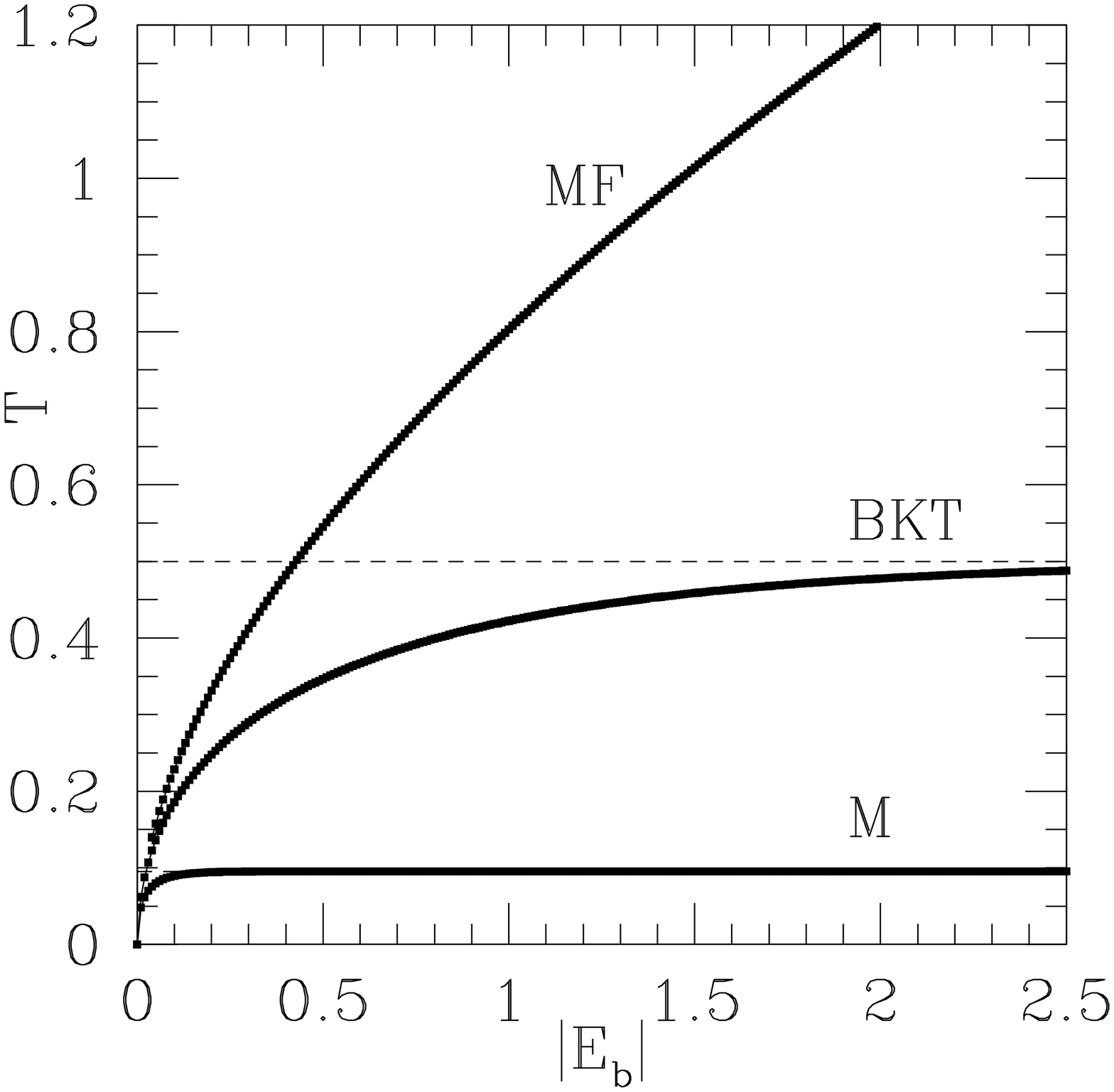}
\includegraphics[width=4.0cm]{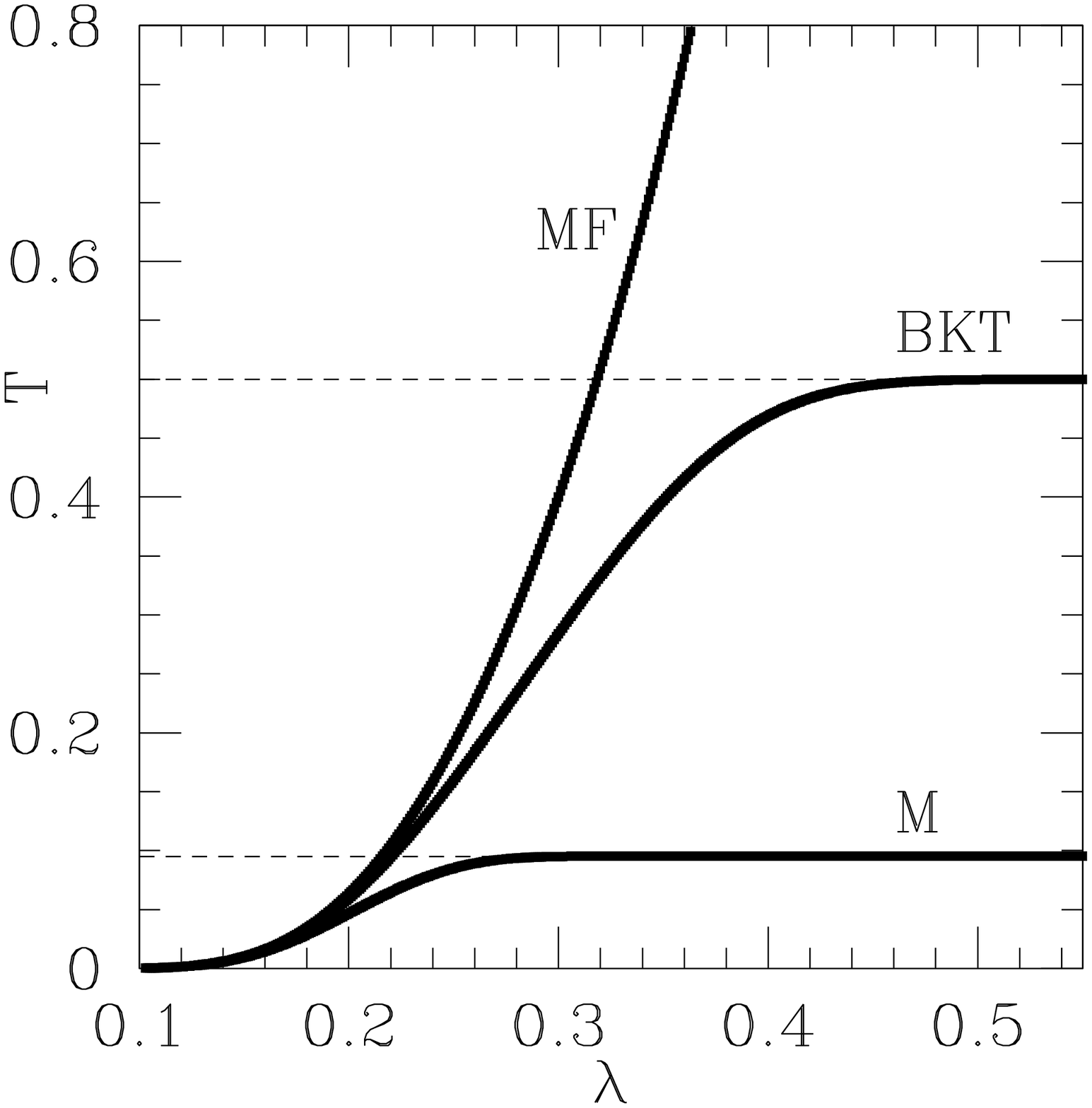}
\end{center}
\vspace{-5mm}
\caption{\small Mean field (MF), Berezinskii-Kosterlitz-Thouless (BKT) and 
vortex-antivortex lattice melting (M) critical temperatures (in units 
of $\epsilon_F$) as functions of (a) absolute value of the binding 
energy $|E_b|$ (in units of $\epsilon_F$) and (b) interaction strength 
$\lambda$ (in units of $g_{\rm 2D}^{-1}$, where $g_{\rm 2D}$ is the 
two-dimensional density of states).}
\label{fig:tc}
\end{figure}

\begin{figure}
\begin{center}
\includegraphics[width=4.2cm]{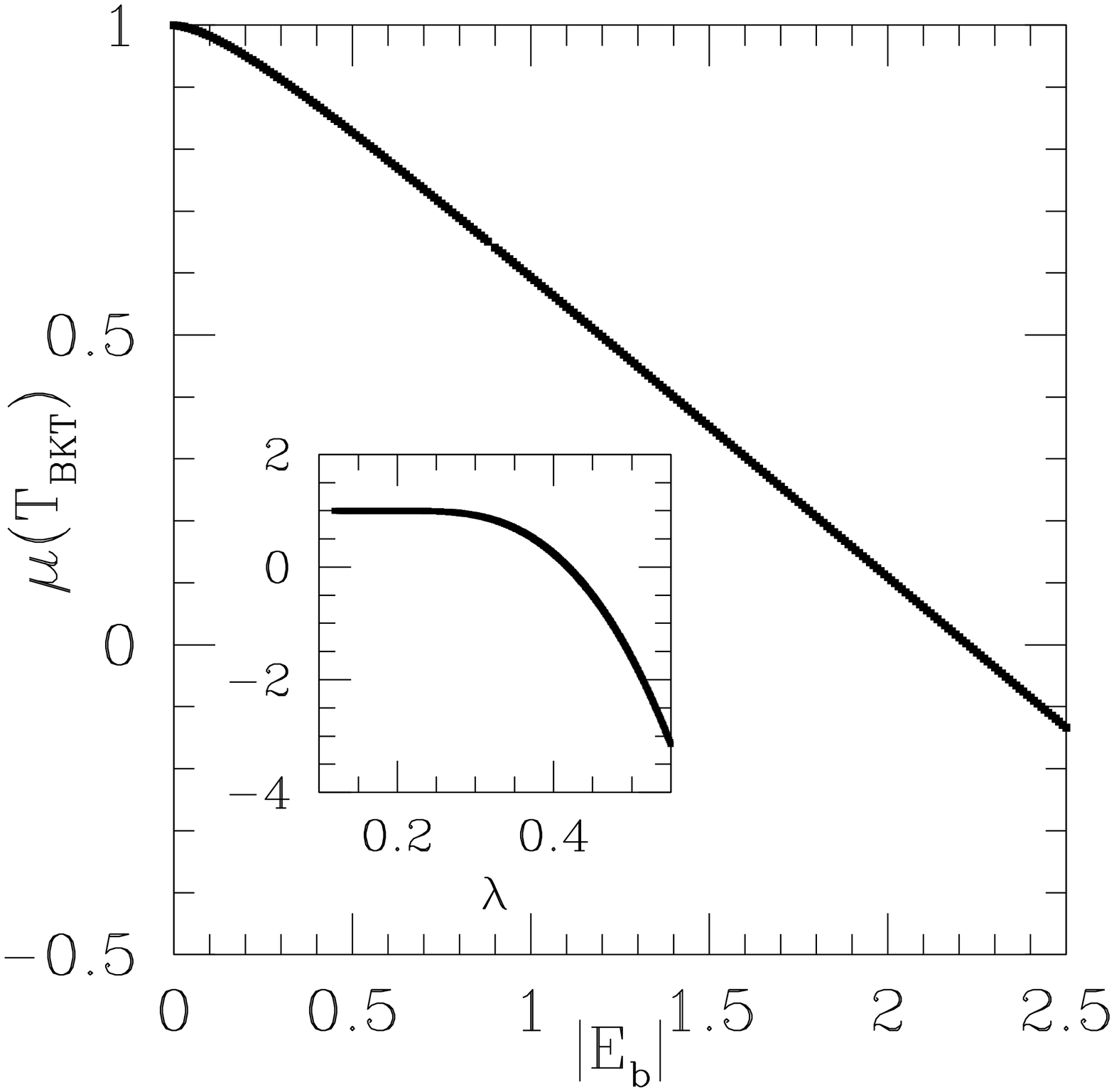}
\includegraphics[width=4.2cm]{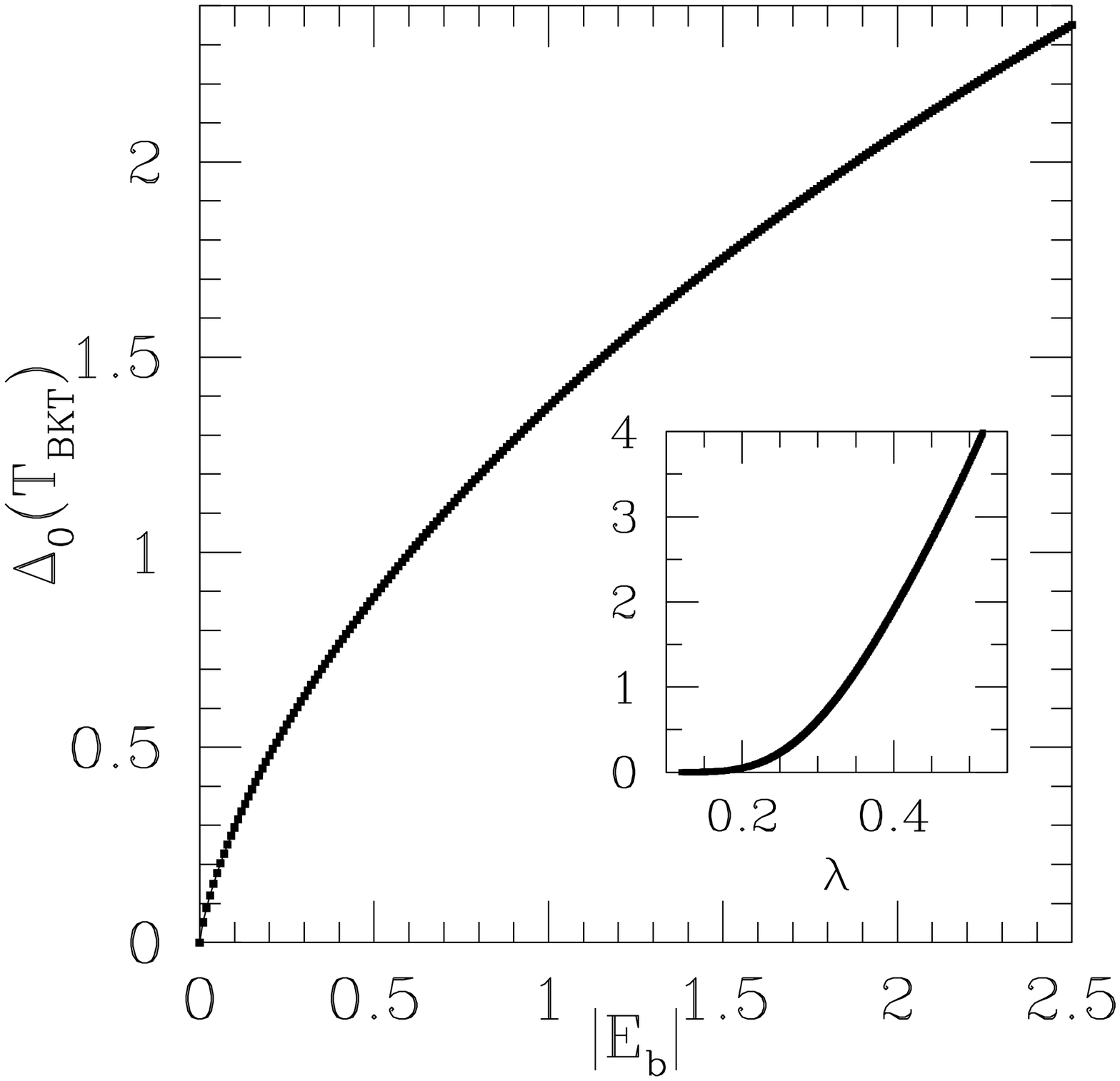}
\end{center}
\vspace{-5mm}
\caption{\small (a) Chemical potential and (b) order parameter amplitude
(both in units of $\epsilon_F$) evaluated at the transition 
temperature $T_{\rm BKT}$, as functions of $|E_b|$ (main figure)
and of the interaction strength $\lambda$ (in units of $g_{\rm 2D}^{-1}$)
(inset).}
\label{fig:mu-delta}
\end{figure}

The equation for the critical temperature $T_c = T_{\rm BKT}$ 
is determined by the Kosterlitz-Thouless~\cite{KT-72} condition
\begin{equation}
\label{eqn:KT}
T_{\rm BKT} = \frac {\pi} {2} \rho_s ( \mu, \Delta_0, T_{\rm BKT} ).
\end{equation}
The self-consistent solutions of Eqs.~(\ref{eqn:op}), (\ref{eqn:number}) 
and (\ref{eqn:KT}) determine $\mu$, $\Delta_0$ and $T_{\rm BKT}$ as functions 
of the two-body binding energy $E_b$ (or interaction strength $\lambda$).
Solutions for $T_{\rm BKT}$ are shown in Fig.~\ref{fig:tc}. 
The curves labeled MF represent the mean-field (saddle-point) solution for
$T_c = T_{\rm MF}$, which is obtained by solving only  
Eqs.~(\ref{eqn:op}) and (\ref{eqn:number}) with $\Delta_0 = 0$.
The disparity between the BKT and MF 
solutions is larger with increasing $\lambda$ (or $|E_b|$), indicating that 
strong coupling results are dramatically affected by phase fluctuations.
Furthermore, $T_{\rm BKT} = 0.5 \epsilon_F$ in the strong coupling limit
where $\rho \sim n/M$, with $n = N_0/L^2$ being the fermion density. 
This shows that although there is no finite $T_c$ for BEC in 2D, there is 
still a superfluid transition with a {\it high} $T_c \propto n$.
Finally, $\mu(T_{\rm BKT})$ and $\Delta_0(T_{\rm BKT})$ are shown in 
Fig.~\ref{fig:mu-delta} as functions of $|E_b|$ or $\lambda$. 
Notice that $\mu(T_{\rm BKT})$ ($\Delta_0(T_{\rm BKT})$) is a monotonically 
decreasing (increasing) function of $|E_b|$ or $\lambda$, and that $\mu$ 
changes sign at $|E_b| \approx 2.23$ ($\lambda \approx 0.419)$.
%

\begin{figure}
\begin{center}
\includegraphics[width=7.5cm]{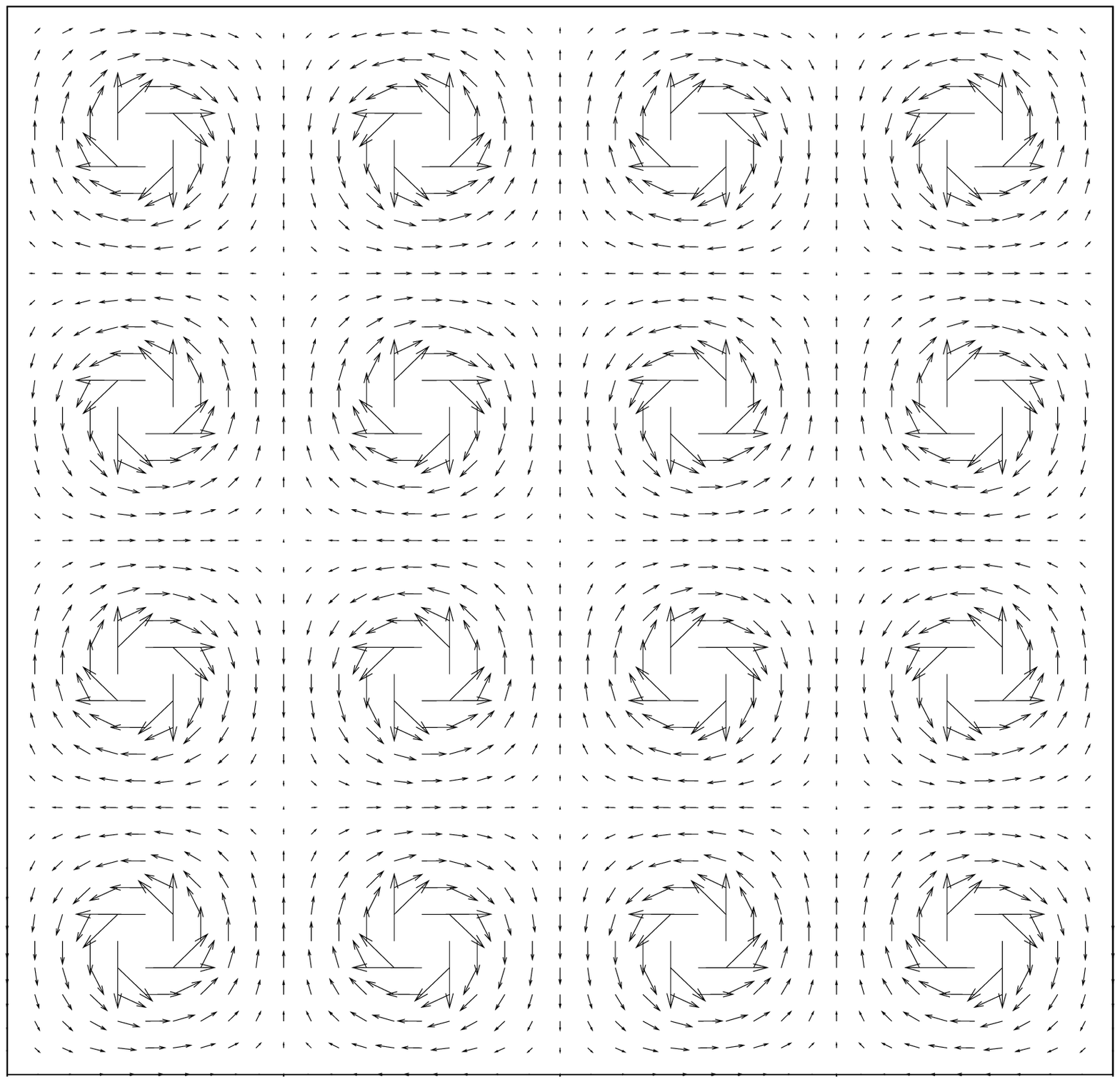}
\end{center}
\vspace{-6mm}
\caption{\small Square vortex-antivortex lattice in the strong coupling
limit.}
\label{fig:va}
\end{figure}

Since the vortex part is directly related to the transverse 
velocity $v_t ({\bf r}) = \nabla \theta_{v} ({\bf r})$,
where $\nabla \cdot v_t ({\bf r}) = 0$, we can express the vortex 
contribution to the action via 
$\nabla \times v_t ({\bf r}) = 2\pi {\bf \hat  z} 
\sum_i n_i \delta ({\bf r} - {\bf r}_i)$, 
where $n_i = \pm 1$ is the vortex topological charge (vorticity). 
Thus, we focus next only on the vortex action, which becomes
$$
S_{v} =  \frac {H_v} {T} = (2\pi) \frac {\rho_s}{2} 
\sum_{i \ne j} n_i n_j G({\bf r}_i - {\bf r}_
j^{\prime})
+ \sum_{i} E_c n_i^2,
$$
where $H_v$ is the vortex Hamiltonian, $E_c$ is the vortex core energy,
and $G({\bf r}_i - {\bf r}_j^{\prime})$ plays the role of the interaction 
potential between topological charges $n_i$ and $n_j$, and satisfies 
$
\nabla^2_{\bf r} G({\bf r} - {\bf r}^{\prime}) = 
\delta ({\bf r} -{\bf r}^\prime).
$
Given that in the vicinity of $T_{\rm BKT}$ VA pairs are formed, 
it is possible that they crystalize into a solid at a lower temperature $T_M$.
We note that the vortex action just obtained allows for the appearance of a 
VA solid (with well defined lattice structure) at a stable minimum of 
the interaction potential. 
At $T \ll T_M$, the lattice configuration that is compatible with
short range core repulsions is such that vortices and antivortices form two 
square sublattices of side $b = a \sqrt{2}$, where $2a$ is the VA pair size. 
This VA lattice can be constructed as a superposition
$
\theta_L (x,y) = 
\sum_{i,j} c_{ij} \theta_{\rm VA} (x - ib, y - jb/2) ,
$
where 
$$
\theta_{\rm VA} (x, y) = 
\arctan \left( \frac {2a y} {a^2 - x^2 - y^2} \right) .
$$
For convenience, we choose the line connecting 
the vortex to the antivortex in a pair to be along the $x$ axis.
The optimal coefficients correspond to $c_{ij} = (-1)^{j}$, where the VA pairs
orient themselves like dipoles in an anti-ferroelectric material. 
All topological dipoles are aligned along the $x$ axis and anti-aligned 
along the $y$ axis, as shown in Fig.~\ref{fig:va}.
However, there is no easy axis as the system is rotationally invariant. 
The configuration where VA pairs are {\it aligned} as dipoles in a 
ferroelectric material is higher in energy.

As the temperature is raised towards $T_M$, the importance of 
vibrations and defects of the VA lattice increases and eventually 
causes its melting into a VA liquid state. In order to obtain
the correct equations of motion, we have to recall that 
vortices and antivortices move perpendicularly 
to the applied force~\cite{fetter-67}. Thus, their dynamics is Eulerian 
instead of Newtonian as in a lattice of atoms. 
This leads to the equation of motion
\begin{equation}
n_i \frac {\partial {\bf r}_{i}} {\partial t} =
\frac {\hbar} {M} {\bf \hat z} \times {\bf F}_i,
\end{equation} 
where 
$
{\bf F}_i = 
-\nabla_{{\bf r}_i} \sum_{j \ne i} n_i n_j G ({\bf r}_i - {\bf r}_j)
$ 
plays the role of the force. 
We can define displacement fields similar to the phonon problem in solids as
${\bf u}_p ({\bf R_i}) = {\bf r}_p ({\bf R}_i) - {\bf R}_i$, 
with the index $p = V, A$ indicating a vortex (V) or an 
antivortex (A) sublattice, and ${\bf R}_i$ labeling the equilibrium lattice 
sites. This equation produces two modes in the long 
wavelength limit: a longitudinal ($L$) mode with frequency 
$\omega_{L} = c_{L} q$, with $c_{L} = ({\hbar/M}) (\gamma_{L}/b)$, 
and a tranverse ($T$) mode with frequency $\omega_{T} = c_{T} q$, with
$c_{T} = (\hbar/M) (\gamma_{T}/b)$, where $\gamma_{L} < \gamma_{T}$ 
are dimensionless constants. The corresponding eigenvectors are
${\bf \eta}_L = {\bf u}_V + {\bf u}_A$, and 
${\bf \eta}_T = {\bf u}_V - {\bf u}_A$.
The harmonic action in diagonal form is
$$
S_{vh} = \frac{\pi \rho_s} {b^2} \int d^2{\bf r} 
\left[
\alpha_{L_1} L_{ij}^2 + \alpha_{L_2} L_{ii}^2
+
\alpha_{T_1} T_{ij}^2 + \alpha_{T_2} T_{ii}^2
\right],
$$
where
$0 < \alpha_{\Lambda_i} < 1$ are dimensionless elastic constants
obtained from the interaction potential, and
$
\Lambda_{ij} = 
\frac {1} {2} [\partial_i \eta_{\Lambda, j} + \partial_j \eta_{\Lambda, i}],
$
with $\Lambda = L, T$.
The VA lattice melting transition 
is then controlled by the Kosterlitz-Thouless-Halperin-Nelson-Young (KTHNY) 
mechanism~\cite{KT-72, nelson-79, young-79}.
The melting temperature $T_M$ is determined by the lowest energy 
mode ($L$) via the relation
\begin{equation}
T_M = \rho_s 
\left[ 
 \frac {\alpha_{L_1} (\alpha_{L_1} + 2\alpha_{L_2})} 
{ 8 (\alpha_{L_1} + \alpha_{L_2}) }
\right] ,
\end{equation}
together with Eqs.~(\ref{eqn:op}) and (\ref{eqn:number}).
The velocities of the VA lattice vibrational modes 
are related to the dimensionless elastic constants
through 
$
\gamma_{L} = \sqrt{\alpha_{L_1}}/2 
$
and
$
\gamma_{T} = \sqrt{ \alpha_{T_1}}/2.
$
The measurement of sound velocities can be useful to extract information 
about the short-range nature of the interaction forces 
through $\alpha_{L_1}$ and $\alpha_{T_1}$.
The values used in Fig.~\ref{fig:tc} correspond to 
$T_M = 0.3 \rho_s (\mu, \Delta_0, T_M)$. 

%


In summary, we have studied the superfluid state of Fermi gases in a 
strongly confining one-dimensional optical lattice that suppresses 
quantum mechanical atom transfer (tunneling),
making the Fermi system two-dimensional.
In the $s$-wave channel, we have found that the Fermi gas undergoes a 
superfluid BKT transition and that vortex-antivortex (VA)
bound states appear below $T_{\rm BKT}$ during the evolution from weak 
to strong coupling. In the strong coupling limit, $T_{\rm BKT}$ is large 
and controlled by the Fermi energy of the atomic system, thus allowing 
for spontaneous VA pairs to be observed directly (without stirring 
the condensate). In addition, we have found that the ground state of 
the system evolves smoothly from weak to strong coupling,
but supports a square VA lattice. Unlike the case of a lattice of 
identical vortices, which is triangular, a square VA lattice is favored, 
because topological charge frustration pushes the triangular VA configuration 
to higher energy.  In addition, we have shown that the VA lattice melts 
via the KTHNY mechanism of dislocation mediated melting. Because the symmetry 
of the VA lattice is square, we expect that the melted lattice will not be 
hexatic, but quartic. The VA lattice might be detected in a trap release experiment 
without the need of rotating the Fermi superfluid. Experimental chances 
of this observation are higher in the strong coupling limit, where the 
melting temperature $T_M < T_{\rm BKT }$ is also controlled by the Fermi energy.
Although we have not yet studied in detail the possible quartic liquid VA 
state, there can be an additional transition into an isotropic VA liquid 
before the BKT transition is reached. 
In case experiments can be done using a velocity sensitive probe instead 
of a density probe as usual, not only the vortex cores would be seen, 
but also the direction of their rotation (vorticity).
Lastly, in addition to the spin wave mode, we have found two VA lattice 
vibrational modes, one longitudinal and one transverse.
All these modes disperse linearly,
but the presence of a transverse mode is also 
a direct indication of the existence of the VA lattice.

\acknowledgments{We thank NSF (DMR 0304380) for financial support.}


\end{document}